# Capacity gain in Li-ion cells with silicon-containing electrodes

Marco-Tulio F. Rodrigues, Charles McDaniel, Stephen E. Trask, Daniel P. Abraham

Chemical Sciences and Engineering Division, Argonne National Laboratory, Lemont, IL, USA

**Contact:** marco@anl.gov

**Abstract**

Silicon-containing lithium-ion batteries can exhibit capacity gain early in life, which makes forecasting future cell behavior difficult. We have observed these anomalous trends even in conditions where known mechanisms, such as overhang equalization and excessive electrolyte oxidation, are unlikely to be significant. Here, we combine simulations and experiments to analyze four cases that can produce increased capacity in Si cells. Three of these pathways relate to "break-in" processes, where improved mass and charge transport can lead to increased access to active electrode domains and decreased cell impedance. The fourth case occurs at high levels of prelithiation, when the positive electrode (PE) is completely replenished with $Li^+$ at the end of cell discharge. We show that the commonality among these mechanisms is that the underlying transformations change the potentials experienced by electrodes at the end of half-cycles, increasing the $Li^+$ inventory available to the cell. A quantitative framework to describe these effects is presented, enabling these ideas to be extended to other battery systems.

**Introduction**

Rechargeable batteries will lose performance with time and use. A dominant degradation pathway is the net consumption of electrons in parasitic reduction reactions to maintain the solid electrolyte interphase (SEI).[1, 2] Additionally, chemo-mechanical processes can lead to deterioration of interfaces and interconnections, rendering entire domains of active material unresponsive to electrochemistry.[3] As a result, cell capacity is generally expected to decay during deployment.

Yet there are many well-documented cases in which cell capacity actually *increases* during early testing. In several instances, such capacity gains reflect enhanced accessibility to the active material. This can happen due to improved "wetting",[4-6] where delayed pore saturation with the liquid electrolyte initially constrains the volume fraction of the electrode that can engage in charge storage. Capacity gains can also result from "break-in" processes, in which early physical transformations in the active material (like particle fracturing) promote electrolyte infiltration and interfacial contact.[7-11] In both these mechanisms, accessing more domains at the positive electrode (PE) expands the pool of available $Li^+$, leading to higher apparent capacity. Increasing the active surface area of electrodes is also expected to decrease impedance. A drop in resistance or, more generally, an improvement in electrode kinetics has also been presented as a source of capacity gain.[12-15]

Capacity increases can also originate from side reactions at the PE particles. Many materials used in high energy cells will experience high electrical potentials in the charged state, which can lead to chemical or electrochemical oxidation of species present in the electrolyte.[16-19] The resulting surplus of electrons in the PE drives uptake of lithium ions from the electrolyte, replenishing some of the $Li^+$ lost to the SEI.[17, 20, 21] Fundamentally, the measured cell capacity reflects a balance between parasitic reduction (electron trapping at the SEI) and parasitic oxidation (electron gain at the PE).[1, 2, 21] Cells tested at particularly high voltages can reach a state where the rate of parasitic oxidation surpasses that of reduction, producing a net increase in measured capacity.[22-24] These gains, however, cause depletion of the electrolyte Li$^+$ inventory and can ultimately compromise long-term cell performance.[23]

Homogenization processes provide an alternative pathway for capacity gain by releasing lithium ions that were temporarily unavailable to the cell. One way that can happen is through overhang equalization. The negative electrode (NE) of commercial cells has a slightly larger area than the PE to accommodate misalignments. Areas of the NE that are not directly opposite the PE coating (i.e., the "overhang") present delayed electrochemical response, often exhibiting a different Li$^+$ content than that displayed by inner portions of the electrode. Gradients in electrical potential between these inner and outer areas of the NE are slowly mitigated by exchange of Li$^+$ between these regions. Thus, cell operation after extended storage at high states of charge (SOC) can result in Li$^+$ influx from the overhang into the electrochemically active NE region, which can then be transferred to the PE as the cell discharges. This sudden access to extra Li$^+$ can offset losses or, at times, produce an increase in cell capacity.[25-30] This mechanism is notably weaker in certain silicon cells, as the voltage hysteresis creates a thermodynamic barrier for overhang equalization.[31]

To complicate matters further, heterogeneous Li$^+$ distribution can also exist within the inner electrode regions, contributing to premature termination of half-cycles and causing apparent capacity loss.[27, 29, 32] Allowing cells to relax can then even out the SOC across the electrodes and restore that capacity.[26, 29] These non-uniformities, however, have been observed to persist in LFP vs graphite systems maintained at intermediate SOCs, where the existence of plateaus in the voltage profile of both electrodes diminishes the thermodynamic driving force for homogenization.[32-34] Subsequent cycling over an expanded voltage range can lead to recovery of as much as ~10% of the initial cell capacity.[33, 34]

Mechanisms such as wetting, break-in and overhang equalization can all be active during the early stages of cell testing, often producing an initial uptick in discharge capacity. More generally, they introduce transient disturbances to capacity trends overlaid to the response produced by SEI growth and other aging processes. Although machine learning methods have shown success in inferring future performance fade using only data acquired from initial periods of testing,[28, 35-39] forecasting errors can increase significantly in the presence of anomalous capacity behavior. This challenge is especially acute for calendar aging experiments, where sparse data density can necessitate disregarding early measurements to prevent bias.[25]

Our team has been particularly interested in studying the calendar aging of Si-based cells, which remains an issue for these systems.[40-44] The experimental evaluation of time-dependent performance loss is intrinsically slow and would benefit from methods that can accelerate the identification of conditions that extend or worsen calendar life.[38, 41, 42, 45] Our initial attempts to use machine learning to predict future aging from a single month of data have shown promise.[44] Nevertheless, many of the pouch cells we have tested exhibit an initial increase in cell capacity (Figure 1), imposing forecasting challenges. The inherent difficulty of extracting fade rates from an upward capacity trajectory motivated our investigation of the specific mechanisms that can trigger such behavior in Si-based cells.

Here, we discuss four such mechanisms. Three of them – impedance drop, increase in active material capacity in the PE and increase in active material capacity in the NE – could emerge from wetting or break-in processes. The fourth mechanism occurs in certain prelithiated cells, which can paradoxically experience capacity gain as a consequence of SEI growth. We explain how capacity gain arises from changes in the electrical potentials of the electrodes at the end of charge or discharge and provide a mathematical framework to quantify this anomalous behavior.

*Experimental and Simulation*

Voltage profiles for NMC532 and silicon electrodes were acquired from tests vs Li metal performed at rates below C/100. Aging was simulated by modifying the electrode profiles before full-cell reconstruction. Impedance decrease was simulated by shifting the voltage profile of the affected electrode to represent a lowered polarization (as in Figure 2a for the case of cell charging). Increases in accessible active material capacity were emulated by elongating the relevant voltage profiles (Figure 2b and 2c); simulating this aging mode also involved small relative shifts between the PE and NE profiles, as discussed in the Appendix. Finally, loss of $Li^+$ inventory (LLI) can be represented through changes in the lateral offset between the voltage profiles of the PE and NE (Figure 2d). We refer the reader to refs. [3, 46] for a detailed discussion about these transformations and how they relate to each aging mode. The simulations used here can approximate cell behavior at slow rates (< C/10), and additional information can be found in refs. [1, 3].

The voltage profiles obtained experimentally from half-cells (Figure 3a) can be combined to construct a full-cell profile. Because cell voltage is the difference between the instantaneous electrical potentials of the PE and NE, the corresponding full-cell voltage profile can be obtained by using Python to perform point-by-point subtraction between the electrode profiles (Figure 3b, solid trace). Thus, the full cell profile in Figure 3b is implicitly contained within the electrode profiles shown in Figure 3a. The termination of each half-cycle occurs when the calculated cell profile reaches pre-specified voltage cutoffs (like 4.2 V in Figure 3b). Note that the x-axes in these charts are in units of Initial Cell SOC. Assuming that Figure 3a and 3b represent a cell after formation, the end of charge (EOC) occurs when the Initial Cell SOC=1. Only portions of the electrode profiles between 0 and 1 are actively utilized within the half-cycle represented in Figure 3a; these portions emulate what would be measured by a hypothetical reference electrode present in the cell.

Figure 3c includes a case where the NE experiences transformations due to aging (dashed orange trace) while the PE remains unchanged; the corresponding full-cell profile is shown as the dashed trace in Figure 3b. This transformation alters the position along the x-axis where the vertical gap between the PE and NE profiles equals the assumed voltage cutoff. Consequently, the simulated aging causes a shift in the EOC (Figure 3b,c, vertical dashed line), indicating changes to the capacity delivered by the cell. In a more general sense, this provides a direct measure of endpoint slippage.[1, 3] Throughout this manuscript, we use representations such as Figure 3c to identify which behaviors can lead to capacity gain, as it allows for better visualization of the assumed changes in PE and NE voltage profiles. Note that the active window of later cycles may extend beyond that bounded by 0 and 1 at the Initial Cell SOC scale (Figure 3b,c).

Experiments to illustrate capacity gain due to LLI were performed in coin cells containing Si electrodes. The ~200 nm particles used in these electrodes were obtained by milling Si boules.[47] The electrode contained 80 wt% Si, 10 wt% C45 carbon (Timcal) and 10 wt% of an aromatic polyimide binder, coated on 10-μm battery-grade copper foil. This electrode was heat-treated at 350 °C for 1 hour before use, as discussed in ref. [43]. The total coating loading was 1.13 mg/cm$^2$ (yielding ~2 mAh/cm$^2$ within 0.05-0.7 V vs Li), the coating thickness was 13 μm and the porosity was 56.2%. The positive electrode comprised 90 wt% NMC532 (Toda), 5 wt% C45 carbon and 5 wt% polyvinylidene fluoride binder (Solvay 5130), coated on 20-μm battery-grade aluminum foil.

The electrode presented coating loading and thickness of 11.35 mg/cm$^2$ and 42 μm, respectively, with a calendered porosity of 33.8%. Both PE and NE in these cells were 14-mm discs, spaced by a 16-mm Celgard 2500. Prelithiation was performed by cycling Si vs Li cells four times at C/10 (0.05-0.7 V) and a final time between 0.05 and 0.55 V. Cells were then disassembled and the harvested electrodes were dipped in dimethyl carbonate to remove electrolyte remnants. The washed electrodes were then paired with NMC811 in fresh cell, which were tested at C/10 between 4.2 V and a designated discharge cutoff (2.5, 3.1 or 3.2 V).

Additional data was acquired from xx3450 single-layer pouch cells, using electrodes with the same composition as discussed in ref. [43] with a NE containing 70 wt% of SiO$_x$ (Osaka Titanium Technologies Co.) and 16 wt% graphite (Hitachi), but with a loading of ~3.1 mAh/cm$^2$ (50-700 mV vs Li). The cells shown in Figure 1 were built without prelithiation, by using a NMC811 PE with sufficiently high capacity (~5.0 mAh/cm$^2$, 2.5-4.3 V vs Li) to compensate for the initial irreversibility of SiO$_x$. In these cells, the PE and NE initially operated within effective potential windows of approximately 3.8-4.25 V and 70-700 mV vs Li/Li$^+$, respectively. Area-specific impedance (ASI) was measured by applying 30-second long 1C current pulses. ASI data from three-electrode cells are presented in more detail in ref [48]. All cells discussed here contained an electrolyte comprising 1.2 M LiPF$_6$ in 3:7 wt ethylene carbonate : ethyl methyl carbonate (Tomiyama), with 3 wt% fluoroethylene carbonate (FEC, Solvay) as additive, with testing performed at 30 °C.

***Results and Discussion***

***Impedance decrease.*** Impedance signatures are generally good indicators of stabilization of early processes in the cell and of pore saturation with electrolyte.[49] In this section, we explore how an initial decrease in impedance affects the capacity of a hypothetical NMC532 vs Si cell.

Figure 4a,b shows the voltage profiles for the PE and NE during charging and discharging, respectively. In this simulation, it is assumed that the impedance of the NE decreases linearly during the initial 30 cycles, causing the electrode polarization to decrease by a total of ~6 mV (Figure 4c). For context, this is the change expected at C/3 for a hypothetical 3 Ah cell with an initial resistance of 30 mΩ (refs. [30, 50]) undergoing a 20% impedance drop. This variation in impedance shifts the NE profile towards higher potentials during charge (Figure 4d) and towards lower potentials during discharge (Figure 4e), thereby displacing the positions of the EOC and of the end of discharge (EOD), as indicated by the vertical lines in both panels.

During charging (Figure 4a,d), the upward movement of the NE profile due to lower polarization results in the voltage cutoff being reached at a higher PE potential; this enables additional PE delithiation and increases the capacity delivered during the charge half-cycle. During discharging (Figure 4b,e), the downward shift of the NE profile forces the voltage cutoff to occur at a lower PE potential, resulting in a higher extent of PE relithiation. Hence, the discharge capacity of the cell increases due to both additional delithiation (Figure 4d) and additional relithiation (Figure 4e) of the PE, leading to a cell capacity increase of ~0.54% over the first 30 cycles (Figure 4f).

We have observed these same qualitative trends in various tests using Si-rich electrodes. Figure 5a shows the impedance measured during calendar aging at 100% SOC and 30 °C of the same cells shown in Figure 1, containing a NMC811 PE and a $SiO_x$-rich NE. Their impedances consistently decrease during the initial 2-3 months of aging, which coincides with the slight capacity increase early in life (Figure 1). Complementary experiments using similar electrodes in a three-electrode configuration indicate that this unusual impedance change originates at the NE (Figure 5b). We speculate that this behavior is associated with post-expansion evolution of the electron-conduction network in the electrode after the initial swelling of $SiO_x$ particles.[48] Although protracted wetting of electrode pores could produce similar outcomes, it is unlikely to be the dominant mechanism here due to the high initial NE porosity (> 60%).

A quantitative discussion on how the discharge capacity of a cell is affected by changes in impedance is provided in the Appendix. Generally, capacity gain due to impedance decrease will depend on cell properties according to

$$capacity\ gain \propto |\Delta R_{drop}| \left( \left| \frac{1}{\frac{dU_{cell,c}}{dq}\big|_{EOC}} \right| + \left| \frac{1}{\frac{dU_{cell,d}}{dq}\big|_{EOD}} \right| \right) \quad (1)$$

where $|\Delta R_{drop}|$ is the magnitude of the impedance decrease, and the derivatives are the slopes of the voltage profiles of the full-cell at the end of charge (left) and discharge (right). This effect is a function of the voltage profile of the cell: the steeper the terminal slopes, the smaller the terms in the bracket, diminishing the impact of impedance changes on cell capacity. A smaller slope means that changes in cell polarization will cause a larger amount of capacity to be exchanged before the cell can settle at a new EOC or EOD. Hence, while the physical origin of the capacity gain is the decrease in cell impedance, the shape of the voltage profiles modulates how these changes will translate into cell capacity. Note that the equation only depends on cell-level properties, and thus consequences would apply whether changes come from PE, NE or both.

We can also use the expression in equation 1 to infer the following: i) NEs with a high Si content are more sensitive to impedance changes than graphite, as the slope of the voltage profile at the EOD tends to be smaller in the former;[2] ii) the effect is more pronounced in NMC532-based cells (Figure 4) than in Ni-rich PEs, as cells with the former will tend to be less steep at the EOC; iii) impedance changes are expected to have negligible influence on the capacity on LFP vs graphite systems at low rates, as the cells show steep voltage variations (i.e., it "polarizes") both at the EOC and EOD, as demonstrated by Dubarry et al.[46]

***Gain of active material capacity.*** Wetting or break-in processes can change the accessibility of electrons and ions to active material particles early in life, leading to an effective increase in accessible electrode capacity. This sudden access to extra sites at the negative and positive electrodes in a NMC532 vs Si cell is illustrated in Figure 6 and Figure 7, respectively. When there is gain of NE capacity (Figure 6c), cell charging is accommodated by a larger number of active sites at the negative electrode. As a result, the EOC will occur at a higher NE potential (Figure 6d), which, just as discussed above for impedance, causes additional delithiation of the PE. Interestingly, this leads to a higher overall amount of $Li^+$ being stored at a larger number of NE active particles, but with each hosting a lower $Li^+$ content. That is, the shallower utilization of the NE is more than compensated by the increased number of active Si particles, producing higher charging capacities. A similar scenario arises for gains in PE capacity. Accessibility to additional PE particles (Figure 7c) will lead to increased utilization of sites in the NE, causing the EOC to occur at lower PE and NE potentials (Figure 7d). Here, the charge capacity increases due to access to a larger population of PE particles, though each particle is delithiated to a smaller extent.

The capacity gains arising from increased accessibility at both PE and NE electrodes can be partially offset by shifts in the EOD towards lower SOC (Figure 6e and Figure 7e). In the case of NE, this occurs due to the extra $Li^+$ consumed as the newly accessed particles "catch up" with the $Li^+$ content of their neighbors. Active NE particles hold a small but finite amount of $Li^+$ at the beginning of charge (BOC),[2] which must be provided to the new, "empty" sites as charging progresses. Given this additional Li$^+$ "trapping", the PE cannot be relithiated to prior levels during cell discharge (Figure 6e). The opposite effect occurs when there is an increase in available PE sites. In this case, the Li$^+$ content at the BOC is higher at the newly accessed particles, and this small excess leads to slightly more PE relithiation at the EOD. Consequently, the PE reaches the EOD at lower potentials, constraining the maximum potential achievable by the NE and thus the extent of Si delithiation. Note the contrast with the impedance behavior discussed above: there, discharge capacity increases due to surpluses during both charge and discharge; here, the discharge capacity only increases because additional capacity had been extracted during cell charging.

Quantitative details are discussed in the Appendix. Improved access to active material particles will affect cell capacity according to

$$capacity\ gain \propto (gain\ of\ NE\ capacity) \left| \frac{\left.\frac{dU_{NE,c}}{dq}\right|_{EOC}}{\left.\frac{dU_{cell,c}}{dq}\right|_{EOC}} \right| \qquad (2)$$

$$capacity\ gain \propto (gain\ of\ PE\ capacity) \left| \frac{\left.\frac{dU_{PE,c}}{dq}\right|_{EOC}}{\left.\frac{dU_{cell,c}}{dq}\right|_{EOC}} \right| \qquad (3)$$

where the derivatives are the slopes of the PE, NE and cell at the EOC. As discussed in the previous section, the slopes modulate how the gains in electrode capacity translate into gains in cell capacity. A key distinction is that equations 2 and 3 depend on the slopes of both the cell and the affected electrode, while equation 1 only relates to the former. The ratios in the equations above can be viewed as an expression of the "limitation" of cell charge, indicating by how much the slope of the voltage profile of the cells at the EOC is coming from the NE (eq. 2) or from the PE (eq. 3).[1-3, 51] Because the EOC typically occurs when the slope of the PE profile is much steeper than that of the NE, the ratio in eq. 3 tends to be much larger than in eq. 2, resulting in stronger measurable capacity gains. Accordingly, a ~3% assumed increase in active material capacity in the PE and NE (Figure 7c and Figure 6c) leads to cell capacity gains of ~2.3% (Figure 7f) and only ~0.35% (Figure 6f), respectively.

Experimental evidence suggests that increased NE capacity also contributes to the behavior described in Figure 1. We have recently shown that the $SiO_x$ particles used in those cells display negligible fracturing even after 500 cycles, with electrochemical testing of electrodes harvested from aged cells indicating a net increase in NE capacity.[48] Experiments vs Li metal suggest that much of this increase occurs during the initial cycles, likely due to rearrangements in the conductive network or enhanced electrolyte access to Si domains following the fracturing of micron-sized $SiO_x$ particles.

Although the dependence of equation 2 on the terminal slope of the voltage profile of the NE suggests that capacity gain is possible for Si-rich electrodes, this mechanism is far less relevant in graphite-based cells. The voltage plateau of graphite at high states of lithiation causes the ratio in eq. 2 to approach zero, minimizing cell-level capacity changes from increased NE accessibility. The negligible changes in NE potential at the EOC in these systems cannot drive significant additional delithiation of the PE, and the overall cell capacity remains nearly invariant.

***Additional silicon amorphization.*** One sub-case of NE capacity gain relates to the progressive amorphization of Si domains. Silicon that is initially crystalline (c-Si) requires a specific range of potentials (100-170 mV vs Li/Li$^+$) before it can commence lithiating.[31, 52, 53] Beyond this initial transition, silicon is irreversibly amorphized (a-Si) and subsequently exhibits its characteristic sloped voltage profile (Figure 8a). Because a-Si can be lithiated at higher potentials than c-Si, increased amorphization will lead to more electrochemical activity at potentials above ~100 mV vs Li/Li$^+$ (Figure 8b). Depending on the selected voltage window, progressive amorphization can lead to gains in electrode capacity in half-cell tests vs Li metal (Figure 8c) during early cycling.[52, 53]

From a cell perspective, this behavior can also be described by equation 2. This amorphization mechanism could occur in response to changes in the PE that drive the NE to progressively lower potentials at the EOC, as illustrated in Figure 7d. Similar behavior has been reported by Friedrich et al., where partial isolation of Si during extended aging led to overlithiation of the remaining electrochemically active domains.[54, 55] In their works, however, gains in cell capacity were not observed, as the effect was offset by concurrent Li$^+$ loss due to SEI growth. Figure 6f suggests that capacity gain from increased access to NE domains is relatively weak and therefore most likely to be observed only if concentrated over a limited number of early cycles.

***Loss of lithium inventory.*** There are interesting cases in which LLI can paradoxically produce gains in cell capacity. In ref. [51], we showed that the changes in cell capacity due to LLI can be quantified by

$$capacity\ change = -LLI \left( \left| \frac{\left|\frac{dU_{PE,c}}{dq}\right|_{EOC}}{\left|\frac{dU_{cell,c}}{dq}\right|_{EOC}} \right| - \left| \frac{\left|\frac{dU_{PE,d}}{dq}\right|_{EOD}}{\left|\frac{dU_{cell,d}}{dq}\right|_{EOD}} \right| \right) \qquad (4)$$

Here, *LLI* accounts for the balance between reduction and oxidation side-reactions that can alter the Li$^+$ that is accessible within a cell. The bracketed quantity has been previously defined as the *information factor*,[51] as it determines how well the measured capacity tracks changes in Li inventory. In equation 4, *LLI* can be negative when the rate of parasitic oxidation is higher than that of reduction; these cases are unusual (though possible, ref. [22-24]) and will not be discussed here. For most practical cells, $LLI > 0$. Under these constraints, an increase in cell capacity (that is, $capacity\ change > 0$) can only occur if the information factor is negative, which requires that

$$capacity\ gain\ condition: \left|\frac{\frac{dU_{PE,d}}{dq}\bigg|_{EOD}}{\frac{dU_{cell,d}}{dq}\bigg|_{EOD}}\right| > \left|\frac{\frac{dU_{PE,c}}{dq}\bigg|_{EOC}}{\frac{dU_{cell,c}}{dq}\bigg|_{EOC}}\right| \quad (5)$$

The inequality expressed in eq. 5 delineates the conditions under which SEI growth can, counterintuitively, lead to capacity gain. For that to happen, the cell discharge must be more "PE-limited" than the charge. Such conditions can occur when cells are prelithiated in excess of what is required to compensate for initial irreversible losses at the NE. In this case, an extra "reservoir" of $Li^+$ remains in the NE that offsets part of the $Li^+$ loss during cell aging, prolonging cell life. This behavior also requires the cell to be discharged to a sufficiently low voltage, such that the half-cycle ends when the PE is fully replenished with $Li^+$.

Figure 9 shows the example of a simulated cell where the NE is prelithiated. In contrast to the previously described cases, this causes the NE voltage profile to extend beyond that of the PE at low SOCs (Figure 9a,b), causing the cell EOD to occur when the PE is completely relithiated (Figure 9b). Application of ~3% of LLI (Figure 9c) depletes $Li^+$ from the cell, leading the NE to reach a lower lithium content at the EOC. This causes the charge half-cycle to terminate at a higher NE potential (Figure 9d), prompting additional delithiation of the PE and shifting the EOC towards higher SOCs. The capacity provided by this extra delithiation is smaller than the amount of $Li^+$ consumed by SEI growth, which is why traditional Si cells exhibit net capacity fade.

The oddity of PE-limited cells is that they offer an additional mechanism that interferes with this behavior. Because the EOD is always determined by full relithiation of the PE (Figure 9b,e), cell capacity is strictly defined by how much $Li^+$ can be extracted from the PE during charge.

Thus, the extra delithiation of the PE will lead to capacity gain. An alternative perspective is that, although the EOC occurs at a lower $Li^+$ content in the NE (Figure 9d), the beginning of charge also occurs with fewer $Li^+$ pre-filling the active Si sites (Figure 9a). That is, the NE will fill less during charging but is also a lot emptier to start with, causing a higher amount of capacity to be exchanged. Ultimately, the cell can exhibit as much as ~0.6% gain in capacity even when ~3% of its initial $Li^+$ inventory is consumed by the SEI (Figure 9f). This anomalous behavior persists until the cell discharge ceases to be PE-limited and equation 5 no longer holds. A further consequence is that the discharge capacity will be higher solely because the charge capacity is increasing. Therefore, cells can operate with coulombic efficiencies $\leq 1$ in all cycles and still display progressively higher discharge capacities. We provide experimental validation of this effect below.

Notwithstanding the discussion above, prelithiation need not always cause the cell to exhibit capacity gain. Recall that capacity gain only happens if the PE is "limiting" the cell discharge more strongly than it does the charge. The example below shows that the same cell can exhibit both behaviors depending on its operating conditions. Figure 10a shows PE and NE voltage profiles during discharge for a hypothetical prelithiated cell. The vertical lines indicate where two different EOD scenarios will occur, assuming a discharge cutoff voltage of either 2.5 V (purple) or 3.1 V (gray). Note that the former occurs when the cathode is completely refilled with $Li^+$, while the latter does not. For both cases, a linear LLI trajectory is imposed such that ~20% of the lithium initially contained within the PE is consumed after ~800 cycles (Figure 10b). Despite the common assumed aging profile, the resulting capacity fade is remarkably different (Figure 10c). The cell discharged to 2.5 V exhibits protracted capacity gain for much of the simulation, reaching a maximum when equation 5 is no longer fulfilled. Capacity only starts decaying once discharge is no longer PE-limited, with the cell displaying a minute net loss of ~2% capacity from the 20% imposed LLI. In contrast, the cell discharged to 3.1 V shows no capacity gain, though the final capacity loss was still only ~10%. Here, the relatively low level of capacity fade is due to the "reservoir effect" discussed in previous publications;[1, 2, 48, 51] silicon is not fully delithiated at the EOD, and a progressive increase in NE potentials due to LLI returns some of that Li$^+$ to the PE, compensating for some of the Li$^+$ lost to the SEI. Note that Figure 10 uses the frame of reference of the initial Li$^+$ content of the PE, rather than that of the cell, to enforce a same absolute level of aging and provide a common capacity scale for comparison.

To demonstrate this mechanism experimentally, we assembled three sets of cells using NMC532 as the PE and a prelithiated Si electrode as the NE (see *Experimental and Simulation* for details about electrodes and prelithiation). All cells were charged to 4.2 V but each set used a different discharge cutoff: 2.5, 3.1 or 3.2 V (these voltages do not compare directly to simulated cases in Figure 10, as the initial state of simulated and experimental cells are not the same). Figure 11a shows how the capacity fade trajectory varied with the discharge cutoff: prolonged capacity gain for 2.5 V; capacity gain and then loss for 3.1 V; and monotonic capacity loss for 3.2 V. As expected from the discussion above, capacity gain can occur for cells operating at a coulombic efficiency < 1 (Figure 11b).

***Conclusions.***

Forecasting aging behavior using early-life test data requires trends that present regular behavior over the life of the cell. Consequently, transient mechanisms operating during early cycles can be problematic for these efforts, as they introduce effects that can distort the evolution of capacity fade and bias lifetime predictions. Many of the cells our team studied over recent years, based on Si nanoparticles or micron-sized $SiO_x$, have displayed a net gain in capacity during the initial ~100 cycles or, in the case of static calendar aging, during the first several months of storage. These observations prompted our interest in identifying the conditions that produce such effects.

We used simulations to show that an initial decrease in cell impedance, or increased access to active material domains at both PE and NE, can result in cell capacity gains in Si-based systems. Such transformations modify the electrode potentials reached at the end of charge or discharge, enabling access to additional $Li^+$ from the PE. This analysis allowed us to identify effects at the NE early in life (impedance drop and access to new domains) as the source of capacity gain in some of our Si-based systems, likely arising from restructuring of the conductive network after the initial volume change.

We also discussed a curious case in prelithiated cells where SEI growth can produce net capacity gain. This behavior emerges from the presence of excess $Li^+$ inventory, causing the PE to be fully relithiated at the end of discharge. As a result, the measurable cell capacity becomes essentially what is extractable from the PE within a single cycle. Progressively shallower NE lithiation due to $Li^+$ consumption will drive the PE EOC potentials upwards, introducing more $Li^+$ into the system and raising cell capacity. Data disclosed by companies on pilot-scale Si cells often show anomalous capacity progression, which we suspect could be linked to this mechanism.

Although our simulations only included Si cells, the framework presented here can be extended to any system. This generality is achieved by leveraging analytical expressions from our previous work that quantitatively describe the capacity fade caused by various aging modes. These equations indicate that the capacity lost (or gained) depends on the slope of the relevant voltage profiles at the end of cell charge and discharge. From these relationships, one can infer that some consequences will be weaker in graphite systems. For example, cells with a prelithiated Si-graphite electrode will likely present constant capacity instead of capacity gain. Additionally, the equations also describe the PE's role on the magnitude of these effects. Materials where the voltage profile are less sloped at the EOC, such as LCO or many NMCs, can exhibit stronger responses, as more $Li^+$ can be released into the cell per increment in potential. In contrast, effects are smaller in LFP, NCA or Ni-rich NMCs, which tend to polarize more strongly as the cell nears the end of charge. Naturally, exact consequences will depend on cell balancing and voltage limits.

Overall, the magnitude of the mechanisms we discuss here can be relatively small. Hence, capacity gain is only visible when these effects are concentrated within a brief period of cell life, typically associated with "break-in" processes where the resulting $Li^+$ surplus can outweigh the $Li^+$ lost to the SEI. Alternatively, they will simply dampen the rate of capacity loss and subtly distort aging trends, just like the parasitic oxidation and overhang equalization processes more widely discussed in the literature.

*Acknowledgements*

This research was supported by the U.S. Department of Energy's Vehicle Technologies Office under the Silicon Consortium Project, directed by Brian Cunningham, Thomas Do, Nicolas Eidson and Carine Steinway, and managed by Anthony Burrell. The submitted manuscript has been created by UChicago Argonne, LLC, Operator of Argonne National Laboratory ("Argonne"). Argonne, a U.S. Department of Energy Office of Science laboratory, is operated under Contract No. DE-AC02-06CH11357. The U.S. Government retains for itself, and others acting on its behalf, a paid-up nonexclusive, irrevocable worldwide license in said article to reproduce, prepare derivative works, distribute copies to the public, and perform publicly and display publicly, by or on behalf of the Government.

*Appendix*

*Definitions.*

- $I\Delta R_k$ is the change in PE polarization under a current $I$ caused by the impedance rise $\Delta R_k$ experienced before the onset of cycle $k$
- $N_0$ is the ratio between the initial capacities of the NE and PE profiles used to simulate/fit cells
- $L_{j,k}$ is the LAM of type $j$ experienced between cycles $k-1$ and $k$, expressed in the SOC scale of the affected electrode
- $x$ is the Li$^+$ content of the NE (measured from 0 to 1)
- $y$ is the state of delithiation of the PE (measured from 0 to 1)

*Sign convention.*

- $I, \left.\dfrac{dU_{cell,c}}{dq}\right|_{EOC}, \left.\dfrac{dU_{PE,c}}{dq}\right|_{EOC}, N_0, x_i, y_i > 0$

- $\left.\dfrac{dU_{cell,d}}{dq}\right|_{EOD}, \left.\dfrac{dU_{PE,d}}{dq}\right|_{EOD} < 0$

The equations in Table A1 express the capacity loss per cycle inflicted by each aging mode, as demonstrated in ref. [3]. These expressions can also be used to identify scenarios in which these same aging modes would instead lead to capacity *gain*, by determining the conditions under which $Q_{loss,k} < 0$. Below, we analyze each case in more detail.

***Impedance decrease.*** Inspection of the sign convention for slopes (see below) indicates that the term in parentheses is strictly $\geq 0$. Hence, a drop in cell impedance ($\Delta R_k < 0$) will always lead to $Q_{loss,k} \leq 0$. The magnitude of the effect will depend on the slopes of the voltage profiles of the cell at the end of charge and discharge. For cells that strongly polarize at the end of both half-cycles (like LFP vs graphite), vertical shifts in the voltage profiles will have little effect on the EOC and EOD.

*Gain of active material capacity.* Table A1 omits losses of capacity in the charged state (LAMliNE and LAMdePE in refs. [3, 46]), as assumption of capacity gain would imply in the generation of lithiated NE particles or delithiated PE particles, which is implausible.

In the equations in Table A1 for active material loss, the terms in parentheses are always $\geq 0$. The ratio between electrode and cell slopes is negative for the NE and positive for the PE. With that in mind, analysis of the equations indicates that $Q_{loss,k}$ and $L_{i,k}$ will have the same sign, and thus electrode capacity gain will translate into cell capacity gain whenever the ratios between derivatives are non-zero. For graphite cells, occurrence of the EOC at a plateau leads to $\frac{dU_{NE,c}}{dq}\Big|_{EOC} \approx 0$, and cell capacity is significantly less sensitive to changes in accessible NE capacity.

In the context of our discussion, $L_{i,k}$ and $-L_{i,k}$ present a fundamental physical distinction. Consider the negative electrode as an example. NE particles will typically hold a finite amount of Li$^+$ even when the cell is nominally fully discharged.[2] Disconnection of NE particles in the discharged state will hence also result in a small amount of Li$^+$ loss, as considered in simulations in refs. [3, 46]. However, capacity gain in the context of the present work involves the sudden accessibility to pristine domains of the active material that are completely empty of Li$^+$. Consequently, the equations in Table A1 do not describe this capacity gain phenomenon with exactitude. Nevertheless, it can be shown that accurate description of capacity gain would only lead to small changes in the terms in parentheses involving $x_i$ and $y_i$, and do not affect the semi-quantitative discussion within the present work.

The points above also required changes to the way we simulate variations in capacity. For active material loss, we utilized in ref. [3] a lateral offset of the relevant voltage profile so that half-cycle endpoints were preserved. This ensured that the Li$^+$ content within the remaining active domains of both electrodes remained unaffected by aging. For example, loss of NE at the discharged state (LAMdeNE) was simulated in such a way that it did not produce changes in the discharge endpoint of the cell, in accordance with the framework developed by Dubarry et al.[46]

To simulate capacity gain, a different type of offset had to be introduced. When the cell is fully discharged, the active NE particles still hold an amount of Li$^+$ equal to $x_{EOD}$. If there is sudden access to "empty" NE particles before charging (as we assume here for capacity again), these new domains need first be filled with $x_{EOD}$ worth of Li$^+$ before the entire NE can proceed lithiating in unison. During the charging in cycle $k$, this "matching" would consume an extra amount of Li$^+$ from the PE equal to $-L_{deNE,k} x_{EOD,k-1}$ (or $-N_0 L_{deNE,k} x_{EOD,k-1}$, in units of PE SOC scale). Simulating this filling of fresh NE portions requires an offset of the NE profile away from that of the PE, like what would be caused by a small amount of LLI. Similarly, sudden access to additional PE capacity can increase the accessible Li$^+$ inventory of the cell, leading to a decrease in the offset between profiles by $L_{liPE,k} y_{EOD,k-1}$.

***Loss of lithium inventory.*** Here, LLI is defined as the net capacity balance between parasitic reduction and oxidation per half-cycle. Hence, $2LLI$ is the Li$^+$ consumption within a full cycle of the cell. This expression assumes that the rates of these side-reactions are constant; more general cases are discussed elsewhere.[1]

Under the assumption that $LLI > 0$, capacity gain happens when the term in parenthesis is negative. Inspecting the sign convention, one can note that both ratios between derivatives are positive, requiring the rightmost ratio to be largest for the cell to gain capacity (see equation 5).

Table A1. Equations describing the measurable capacity fade caused by each aging mode, as discussed in ref. [3].

| Aging mode | Cell capacity loss at cycle $k$ ($Q_{loss,k}$) [a] |
|---|---|
| Impedance rise | $I \Delta R_k \left( \dfrac{1}{\left.\dfrac{dU_{cell,c}}{dq}\right|_{EOC}} - \dfrac{1}{\left.\dfrac{dU_{cell,d}}{dq}\right|_{EOD}} \right)$ |
| Loss of NE capacity (delithiated) | $-N_0 L_{deNE,k}(x_{EOC,k-1} - x_{EOD,k-1}) \dfrac{\left.\dfrac{dU_{NE,c}}{dq}\right|_{EOC}}{\left.\dfrac{dU_{cell,c}}{dq}\right|_{EOC}}$ |
| Loss of PE capacity (lithiated) | $L_{liPE,k}(y_{EOC,k-1} - y_{EOD,k-1}) \dfrac{\left.\dfrac{dU_{PE,c}}{dq}\right|_{EOC}}{\left.\dfrac{dU_{cell,c}}{dq}\right|_{EOC}}$ |
| Loss of Li$^+$ inventory (LLI) [b] | $2 LLI_k \left( \dfrac{\left.\dfrac{dU_{PE,c}}{dq}\right|_{EOC}}{\left.\dfrac{dU_{cell,c}}{dq}\right|_{EOC}} - \dfrac{\left.\dfrac{dU_{PE,d}}{dq}\right|_{EOD}}{\left.\dfrac{dU_{cell,d}}{dq}\right|_{EOD}} \right)$ |

[a] Equations valid assuming capacities are in the PE SOC scale; see ref. [3] for discussion on changes in the frame of reference; [b] LLI is measured per half-cycle and assumed to be constant during the charge and discharge of a same cycle.

*References.*

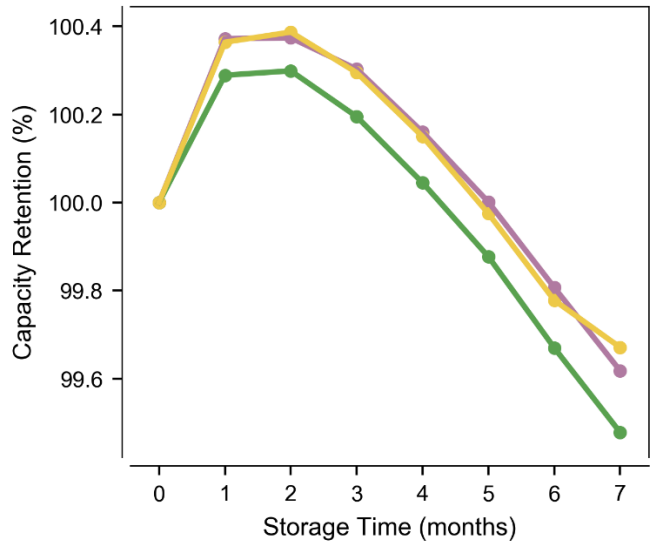

Figure 1. Calendar aging of triplicate non-prelithiated NMC811 vs SiO$_x$ pouch cells stored at 30 °C and 100% state of charge. Cells consistently show capacity gain during the initial reference performance tests, rendering the early data points less reliable for life forecasting.

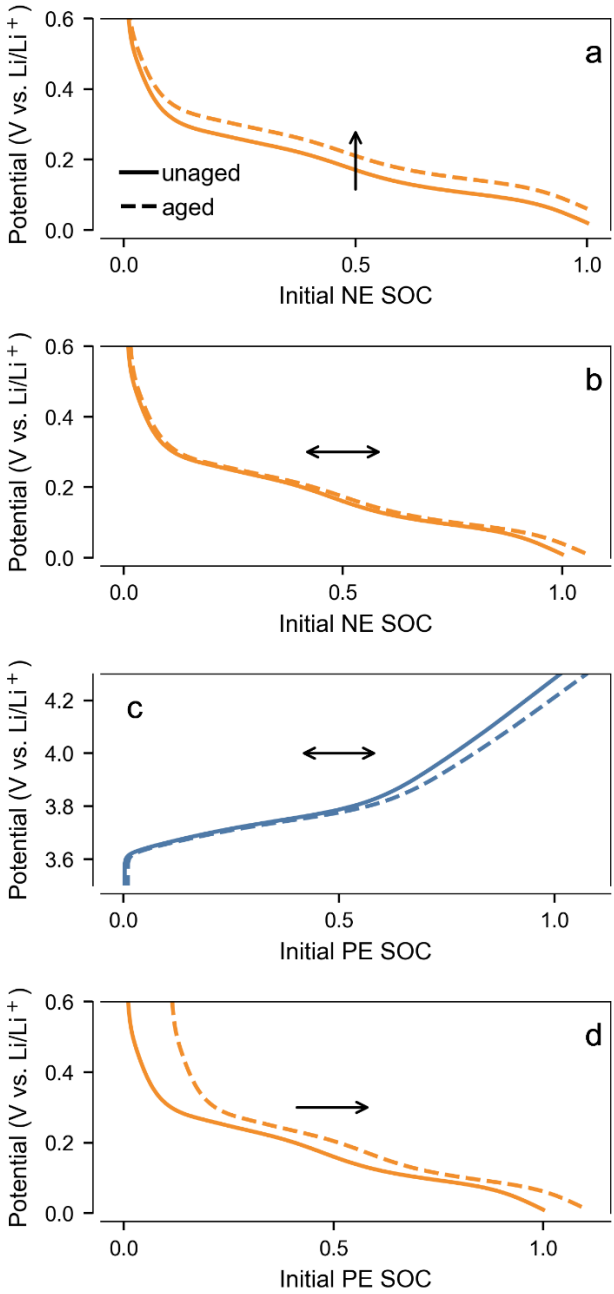

Figure 2. Transformations of voltage profiles used to simulate aging. (a) Impedance decrease. (b) Gain of active material capacity in the negative electrode. (c) Gain of active material capacity in the positive electrode. (d) Loss of Li$^+$ inventory. The x-axis in panels a-c are in the scale of the post-formation Li$^+$ content of the affected electrode; in panel d, the scale of the PE is used to provide a static frame of reference. The legend in panel a applies to all panels.

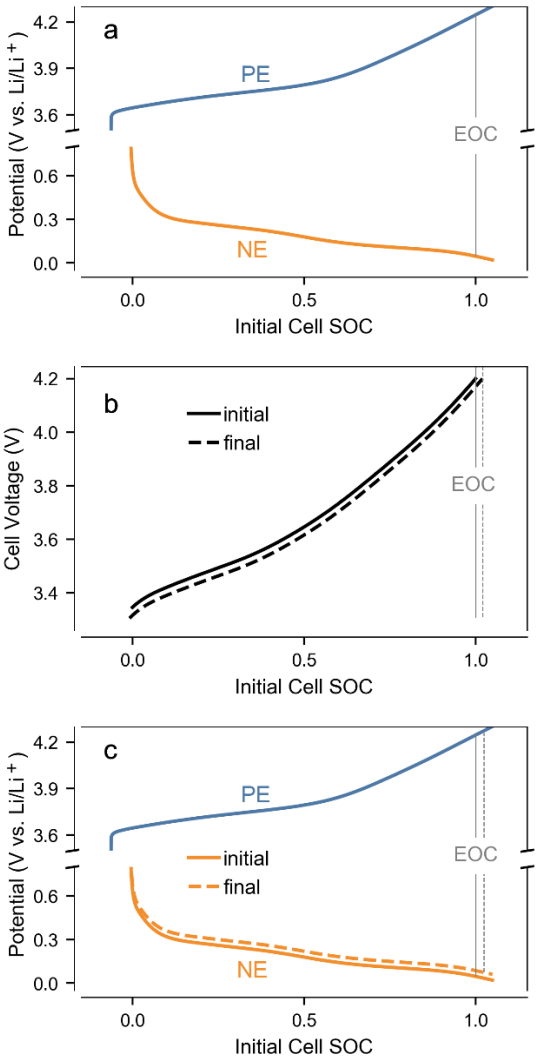

Figure 3. Representing full-cells using half-cell data. (a) Voltage profiles from half-cells can be aligned to indicate the response of electrodes within a cell. Portions of the profiles between 0 and 1 SOC emulate what would be sensed by a reference electrode. (b) Voltage profiles for a hypothetical full-cell before (solid) and after (dashed) simulated "aging" that results in capacity gain. Note the vertical line indicating a shift of the end of charge. (c) Representation of full-cells in panel b using only voltage profiles for half-cells. All dashed lines indicate states after the assumed aging.

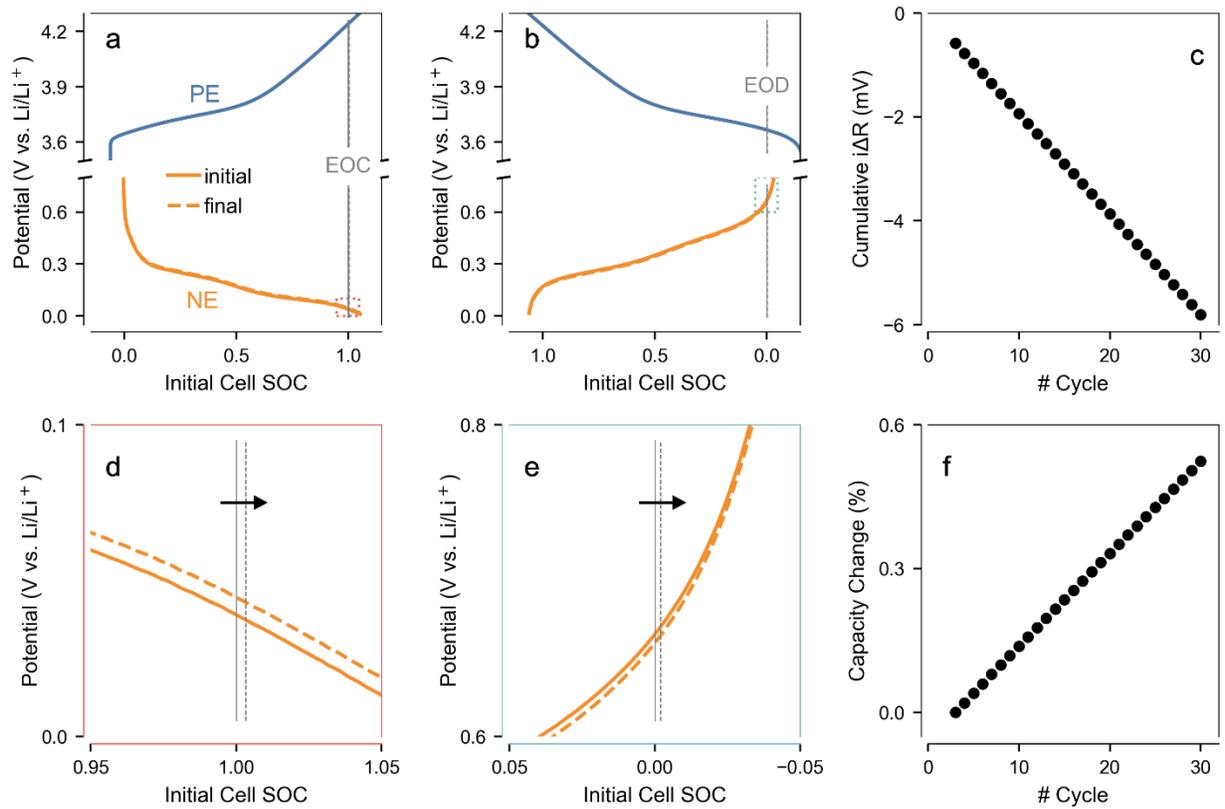

Figure 4. Simulating the effects of a decrease in the impedance of the negative electrode. (a) Electrode profiles during cell charge. (b) Electrode profiles during cell discharge. (c) Cumulative decrease in polarization caused by the assumed impedance drop. (d) Zoomed-in view of panel a around the EOC. (e) Zoomed-in view of panel b around the EOD. (f) Cumulative changes in cell capacity.

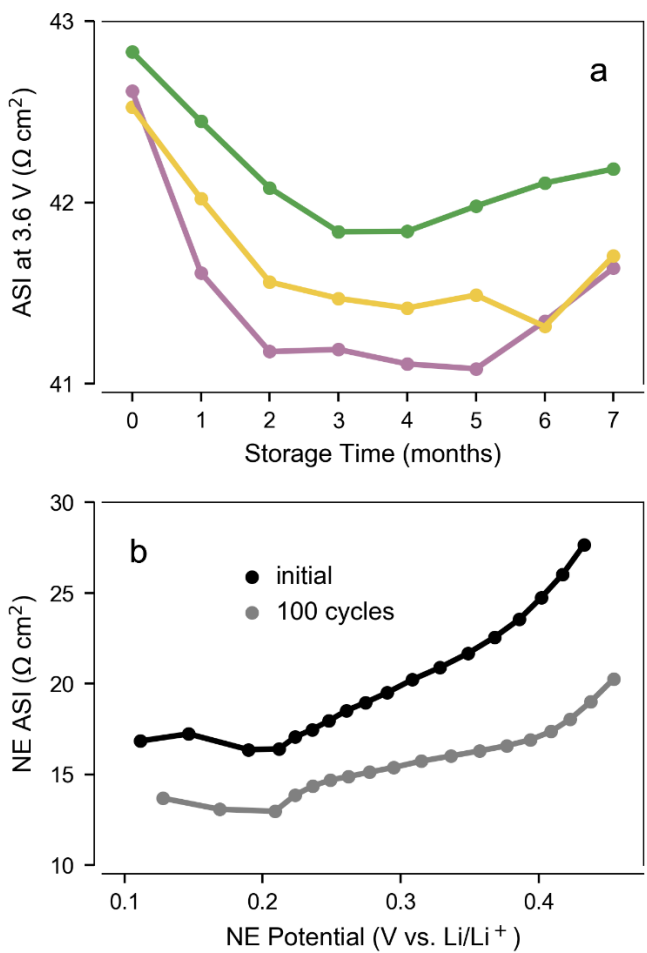

Figure 5. Experimental evidence of impedance drop in $SiO_x$-rich electrodes. (a) Area-specific impedance of cells shown in Figure 1. The decrease in ASI during initial months of testing coincides with the observed capacity gain. (b) ASI data for the NE measured in a 3-electrode cell during cycling of NMC811 vs $SiO_x$. Impedance lowers after initial aging of the cell, suggesting that the anomaly in panel a originates at the NE.

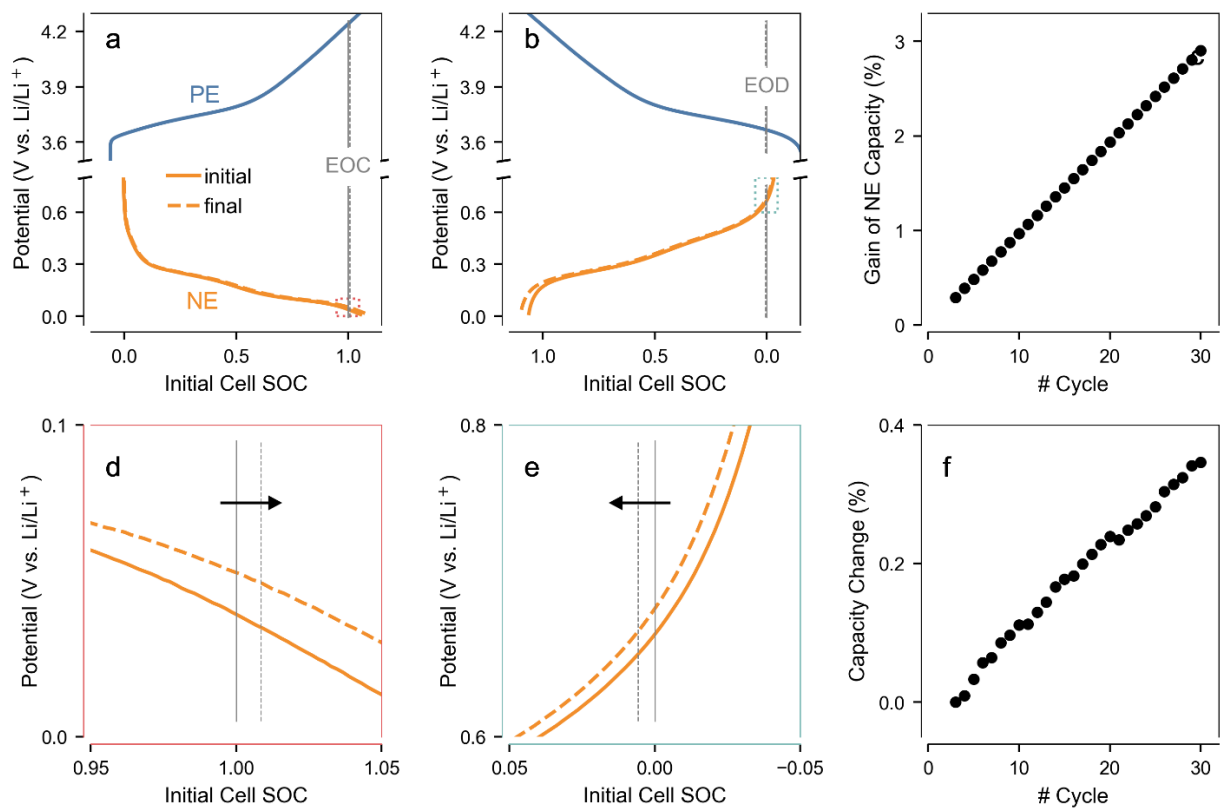

Figure 6. Simulating an increased access to negative electrode capacity. (a) Electrode profiles during cell charge. (b) Electrode profiles during cell discharge. (c) Cumulative gain in NE capacity. (d) Zoomed-in view of panel a around the EOC. (e) Zoomed-in view of panel b around the EOD. (f) Cumulative changes in cell capacity.

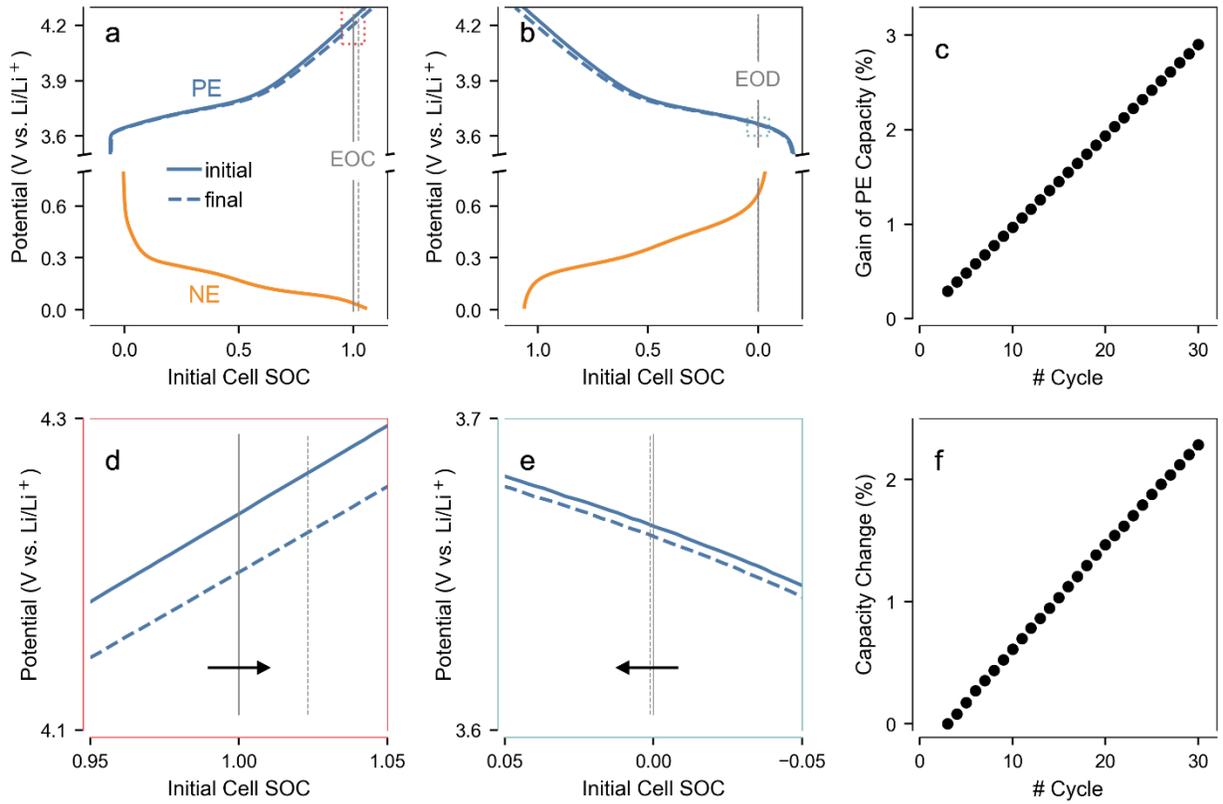

Figure 7. Simulating an increased access to positive electrode capacity. (a) Electrode profiles during cell charge. (b) Electrode profiles during cell discharge. (c) Cumulative gain in PE capacity. (d) Zoomed-in view of panel a around the EOC. (e) Zoomed-in view of panel b around the EOD. (f) Cumulative changes in cell capacity.

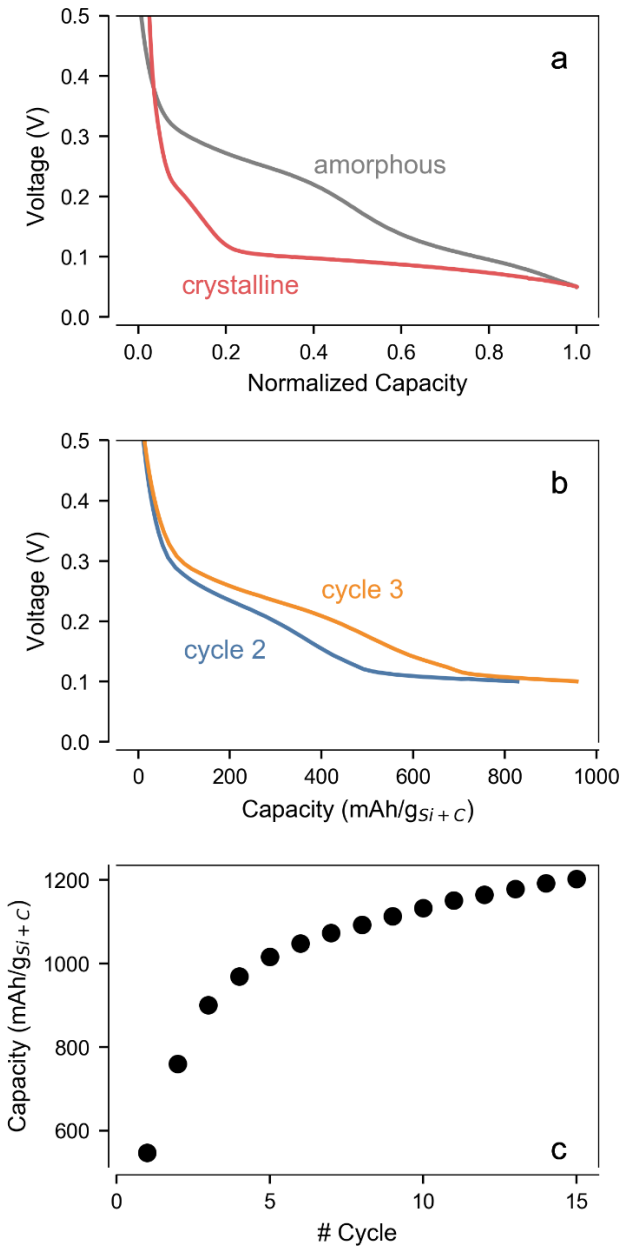

Figure 8. Progressive amorphization of silicon. (a) Voltage profiles for crystalline and amorphous silicon. (b) Profiles from a Si vs Li cell cycled with a 100 mV cutoff. Amorphization of silicon is incomplete at the end of each lithiation, leading to progressive transformation of the active material. (c) Capacity measured in the cell shown in panel b. Capacity gain happens due to the increasing amount of amorphous Si, which can be lithiated at higher potentials.

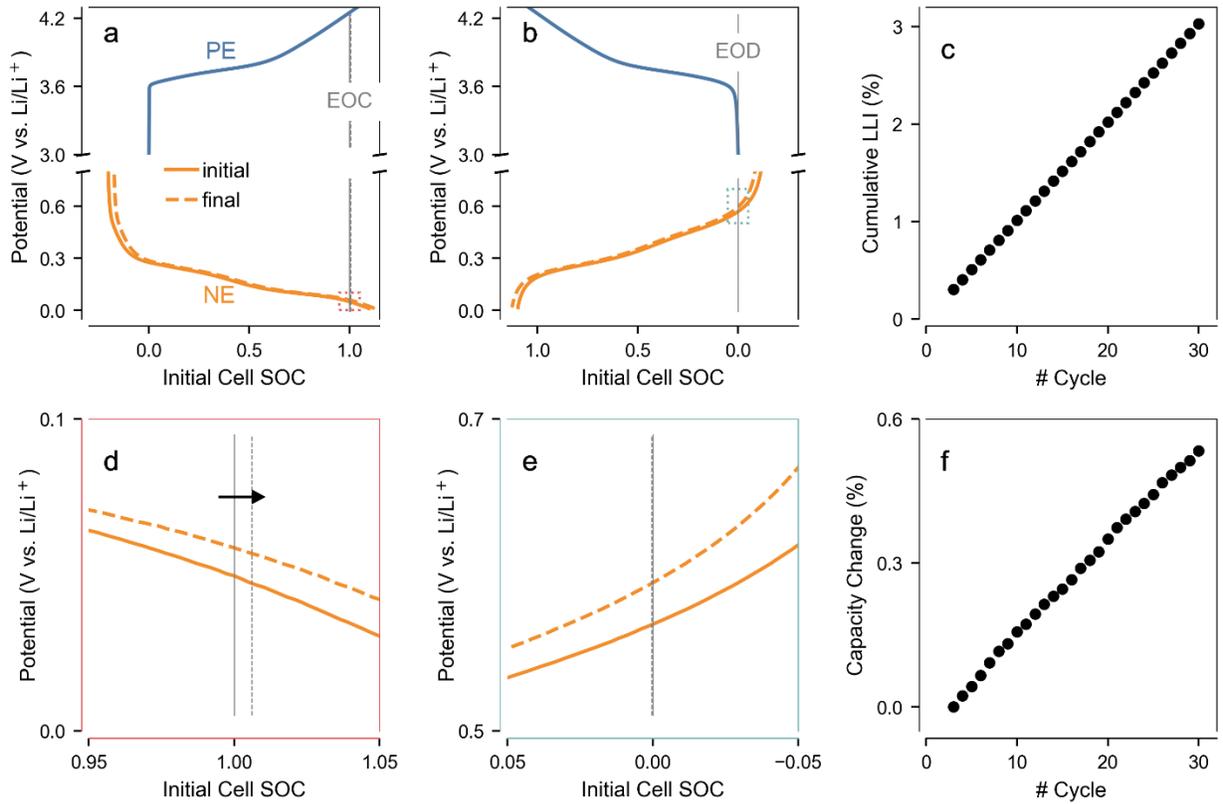

Figure 9. Simulating loss of Li$^+$ inventory in a prelithiated cell. (a) Electrode profiles during cell charge. (b) Electrode profiles during cell discharge. (c) Cumulative loss of Li$^+$ inventory. (d) Zoomed-in view of panel a around the EOC. (e) Zoomed-in view of panel b around the EOD. (f) Cumulative changes in cell capacity.

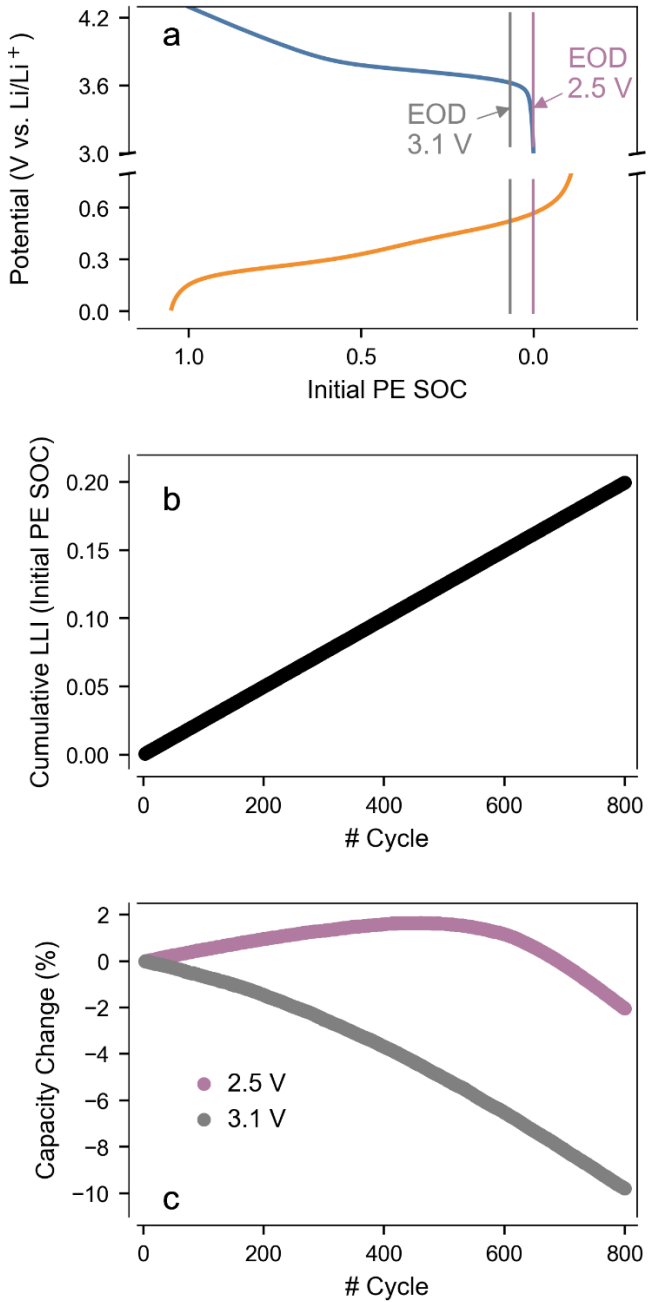

Figure 10. Contrasting scenarios for prelithiated cells. (a) Electrode profiles during discharge, with vertical lines indicating the EOD at two different assumed voltage cutoffs. (b) Cumulative LLI applied to both cells showed in panel a. (c) Capacity change due to LLI. All capacity values are presented with respect to the PE SOC scale to provide a common frame of reference.

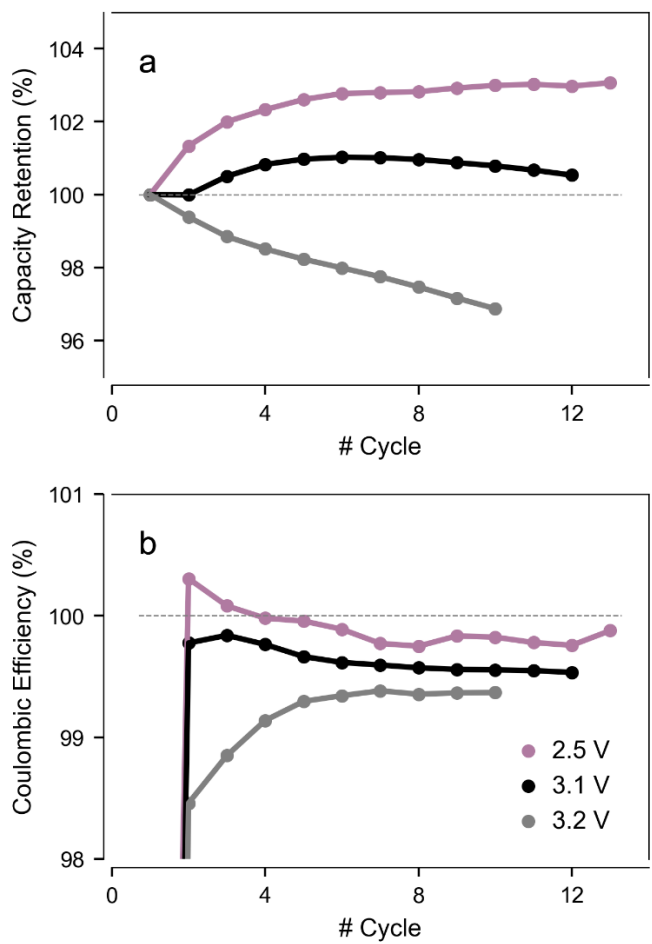

Figure 11. Experimental demonstration of capacity gain in prelithiated NMC811 vs Si cells. (a) Measured capacity. (b) Coulombic efficiency. As predicted in Figure 10, the initial capacity trajectory depends on the selected discharge cutoff, and gains can occur with an efficiency lower than 1.